\documentclass[pra,twocolumn,showpacs,superscriptaddress,amsfonts,amsmath,floatfix]{revtex4}

\usepackage{graphicx}
\usepackage{color}

\begin{document}

%=======================================================================================
\title{Plasmonics in graphene at infra-red frequencies}

\author{Marinko Jablan}
\email{mjablan@phy.hr}
\affiliation{Department of Physics, University of Zagreb, Bijeni\v cka c. 32, 10000 Zagreb, Croatia}

\author{Hrvoje Buljan}
\email{hbuljan@phy.hr}
\affiliation{Department of Physics, University of Zagreb, Bijeni\v cka c. 32, 10000 Zagreb, Croatia}

\author{Marin Solja\v ci\' c}
\email{soljacic@mit.edu}
\affiliation{Department of Physics, Massachusetts Institute of Technology, 77 Massachusetts Avenue, Cambridge MA 02139, USA}

\date{\today}

\begin{abstract}
We point out that plasmons in doped graphene simultaneously enable 
low-losses and significant wave localization for frequencies below that 
of the optical phonon branch $\hbar\omega_{Oph}\approx 0.2$~eV. 
Large plasmon losses occur in the interband regime (via excitation of electron-hole 
pairs), which can be pushed towards higher frequencies for higher doping values. 
For sufficiently large dopings, there is a bandwidth of frequencies from 
$\omega_{Oph}$ up to the interband threshold, where a plasmon decay channel 
via emission of an optical phonon together with an electron-hole pair is nonegligible. 
The calculation of losses is performed within the framework of a 
random-phase approximation and number conserving relaxation-time 
approximation.
The measured DC relaxation-time serves as an input parameter 
characterizing collisions with impurities, whereas the contribution from 
optical phonons is estimated from the influence of the electron-phonon coupling 
on the optical conductivity. 
Optical properties of plasmons in graphene are in many relevant aspects 
similar to optical properties of surface plasmons propagating on 
dielectric-metal interface, which have been drawing 
a lot of interest lately because of their importance for nanophotonics. 
Therefore, the fact that plasmons in graphene could have low losses for certain 
frequencies makes them potentially interesting for nanophotonic applications. 
\end{abstract}

\maketitle
%\narrowtext
%\newpage

%=======================================================================================
\section{Introduction}
\label{sec:intro}

In recent years, an enormous interest has been surrounding the field of 
plasmonics, because of the variety of tremendously exciting and novel phenomena 
it could enable. On one hand, plasmonics seems to be the only viable path 
towards realization of nanophotonics: control of light at scales substantially 
smaller than the wavelength \cite{Barnes2003,Atwater2005,Yablonovitch2005,
Aristeidis2005}. On the other hand, plasmonics is a 
crucial ingredient for implementation of most metamaterials, and thereby 
all the exciting phenomena that they support \cite{Veselago1968,Shalaev2007,
Pendry2000,Smith2004}, including negative refraction, superlensing, and cloaking. 
However, there is one large and so far insurmountable obstacle towards 
achieving this great vision: plasmonic materials (most notably metals) have 
enormous losses in the frequency regimes of interest. 
This greatly motivates us to explore plasmons and their losses in 
a newly available material with unique properties: 
graphene \cite{Novoselov2004,Novoselov2005a,
Novoselov2005b,Zhang2005,Geim2007,Berger2004,Berger2006}. 

Graphene is a single two-dimensional (2D) plane of carbon atoms arranged 
in a honeycomb lattice, which has only recently been demonstrated 
in high quality samples and with superior mobilities 
\cite{Novoselov2004,Novoselov2005a,Novoselov2005b,Zhang2005,Geim2007,Berger2004,Berger2006}. 
This material is a zero-gap semiconductor, which can be doped to high values of 
electron or hole concentrations by applying voltage externally 
\cite{Novoselov2004}, much like in field effect transistors (FET). 
While this kind of control over electrical properties of materials is at the 
heart of modern electronics, it was also demonstrated that the same procedure 
(electric gating) \cite{Wang2008,Basov2008} leads to a dramatic change in 
optical properties of graphene because of its impact on the strong interband 
transitions. Collective excitations (plasmons) in graphene hold potential for 
technological applications as well \cite{Vafek2006,Ryzhii2007a,Rana2008,Wunsch2006,Hwang2007,
Mikhailov2007,Falkovsky2007,Bostowick2007,Ryzhii2007b,Gangadharaiah2008}; 
for example, coherent terahertz sources based 
on plasmon amplification were suggested and discussed in Refs. \cite{Ryzhii2007a,Rana2008}. 
Graphene was predicted to support a transverse electric (TE) mode 
\cite{Mikhailov2007}, which is not present in usual 2D systems with parabolic 
electron dispersion. Thermo-plasma polaritons in graphene have been 
discussed in Ref. \cite{Vafek2006}, pointing out at new opportunities in 
the field of plasmonics. 

\begin{figure*}[!ht]
\centerline{
%\mbox{\includegraphics[width=0.85\textwidth]{./Figures/Figure1.eps}}
%\mbox{\includegraphics[width=0.85\textwidth]{./Figures/Figure1pdf.eps}}
%\mbox{\includegraphics[width=0.85\textwidth]{./Shito/Figure1.eps}}
\mbox{\includegraphics[width=0.85\textwidth]{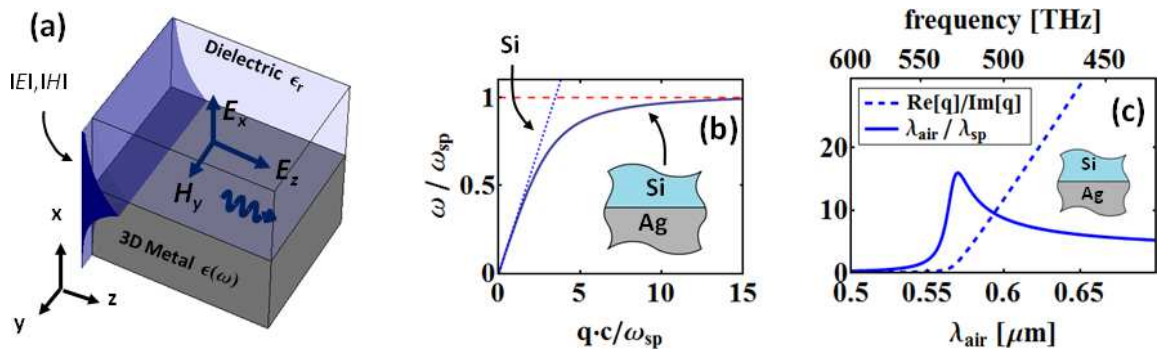}}
}
\caption{
(color online) 
(a) Schematic description of a surface plasmon (SP) on metal-dielectric 
interface. (b) SP dispersion curve (solid blue line) for Ag-Si interfaces; 
dotted blue is the light line in Si; dashed red line denotes the SP resonance. 
(c) Wave localization and propagation length for SPs at Ag-Si interface 
(experimental Ag losses are taken into account).
}
\label{figsp}
\end{figure*}

Here we investigate plasmons in doped graphene and demonstrate that they 
simultaneously enable low-losses and significant wave localization 
for frequencies of the light smaller than the optical phonon 
frequency $\hbar\omega_{Oph}\approx 0.2$~eV \cite{phonondisp}. 
Interband losses via emission of electron-hole pairs (1$^{\textrm{st}}$ order process) can be 
blocked by sufficiently increasing the doping level, which pushes the 
interband threshold frequency $\omega_{inter}$ towards higher values 
(already experimentally achieved doping levels can 
push it even up to near infrared frequencies). 
The plasmon decay channel via emission of an optical phonon 
together with an electron-hole pair (2$^{\textrm{nd}}$ order process) is inactive for 
$\omega<\omega_{Oph}$ (due to energy conservation), however, for frequencies larger than $\omega_{Oph}$ 
this decay channel is nonnegligible. This is particularly important 
for large enough doping values when the interband threshold $\omega_{inter}$ is 
above $\omega_{Oph}$: in the interval $\omega_{Oph}<\omega<\omega_{inter}$ 
the 1$^{\textrm{st}}$ order process is suppressed, but the phonon decay channel is open. 
In this article, the calculation of losses is performed within the 
framework of a random-phase approximation (RPA) and number conserving 
relaxation-time approximation \cite{Mermin1970}; the measured 
DC relaxation-time from Ref. \cite{Novoselov2004} serves as an input parameter 
characterizing collisions with impurities, 
whereas the optical phonon relaxation times are estimated  
from the influence of the electron-phonon coupling \cite{Park2007} on the 
optical conductivity \cite{Stauber2008}.

In Sec. \ref{sec:sps}, we provide a brief review of conventional surface plasmons 
and their relevance for nanophotonics. In Sec. \ref{sec:res} we discuss the 
trade off between plasmon losses and wave localization in doped graphene, 
as well as the optical properties of these plasmons. 
We conclude and provide an outlook in Sec. \ref{sec:conc}.

\section{Surface plasmons}
\label{sec:sps}

Surface plasmons (SPs) are electromagnetic (EM) waves that propagate along 
the boundary surface of a metal and a dielectric [see Fig. \ref{figsp}(a)]; these are 
transverse magnetic (TM) modes accompanied by collective oscillations of 
surface charges, which decay exponentially in the transverse directions 
(see, e.g., Refs. \cite{Barnes2003,Atwater2005} and Refs. therein). Their 
dispersion curve is given by:
\begin{equation}
q_{sp}=\frac{\omega}{c} 
\sqrt{\frac{\epsilon_r \epsilon(\omega)}{\epsilon_r+\epsilon(\omega)}}
\label{convP}
\end{equation}
[see Fig. \ref{figsp}(b)]; 
note that close to the SP resonance ($\omega=\omega_{SP}$), 
the SP wave vector [solid blue line in Fig. \ref{figsp}(b)] is much 
larger than the wave vector of the same frequency excitation in 
the bulk dielectric [dotted blue line in Fig. \ref{figsp}(b)]. 
As a result, a localized SP wave packet can be much smaller than a same 
frequency wave packet in a dielectric. Moreover, this "shrinkage" is accompanied 
by a large transverse localization of the plasmonic modes. These features are 
considered very promising for enabling nano-photonics 
\cite{Barnes2003,Atwater2005,Yablonovitch2005,Aristeidis2005}, 
as well as high field localization and enhancement. 
A necessary condition for the existence of SPs is $\epsilon(\omega)<-\epsilon_r$ 
(i.e., $\epsilon(\omega)$ is negative), which is why metals are usually used. 
However, SPs in metals are known to have small propagation lengths, which are 
conveniently quantified (in terms of the SP wavelength) with the ratio
$\Re q_{sp}/\Im q_{sp}$; this quantity is a measure of how many SP wavelengths 
can an SP propagate before it loses most of its energy. 
The wave localization (or wave "shrinkage") is quantified as 
$\lambda_{air}/\lambda_{sp}$, where $\lambda_{air}=2\pi c/\omega$ (the 
wavelength in air). These quantities are plotted in Fig. \ref{figsp}(c) for 
the case of Ag-Si interface, by using experimental data (see \cite{Yablonovitch2005} 
and references therein) to model silver (metal with the lowest losses for the frequencies 
of interest). Near the SP resonance, wave localization reaches its peak; 
however, losses are very high there resulting in a small propagation length 
$l\approx 0.1 \lambda_{sp}\approx 5$nm. At higher 
wavelengths one can achieve low losses but at the expense of poor wave 
localization.

\section{Plasmons and their losses in doped graphene}
\label{sec:res}

Graphene behaves as an essentially 2D electronic system. In the absence of 
doping, conduction and valence bands meet at a point (called Dirac point) 
which is also the position of the Fermi energy. The band structure, calculated in the tight 
binding approximation is shown in Fig. 2(b) (see Ref. \cite{Bostowick2007} and 
references therein); for low energies the dispersion around the Dirac point 
can be expressed as $E_{n,{\bf k}}= n v_F \hbar |{\bf k}|$, where the 
Fermi velocity is $v_F=10^6$m/s, $n=1$ for conduction, and $n=-1$ for the 
valence band. Recent experiments \cite{Nair2008} have shown that this linear dispersion 
relation is still valid even up to the energies (frequencies) of visible 
light, which includes the regime we are interested in. 

\begin{figure*}[!ht]
\centerline{
%\mbox{\includegraphics[width=0.85\textwidth]{./Figures/Figure2.eps}}
%\mbox{\includegraphics[width=0.85\textwidth]{./Figures/Figure2pdf.eps}}
%\mbox{\includegraphics[width=0.85\textwidth]{./Shito/Figure2.eps}}
\mbox{\includegraphics[width=0.85\textwidth]{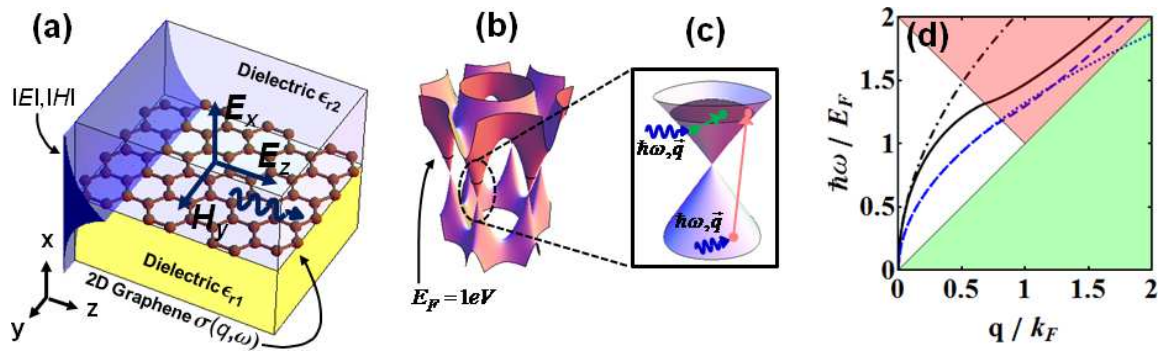}}
}
\caption{
(color online) 
(a) Schematic of the graphene system and TM plasmon modes. Note that the profile of 
the fields looks the same as the fields of an SP [Fig. \ref{figsp}(a)]. 
(b) Electronic band structure of graphene; to indicate the vertical scale 
we show the Fermi energy level for the case $E_F=1$~eV. 
(c) Sketch of the intraband (green arrows) and interband (red arrows) 
single particle excitations that can lead to large losses; these losses 
can be avoided by implementing a sufficiently high doping. 
(d) Plasmon RPA and semiclassical dispersion curves.
Black solid (RPA) and black dot-dashed (semiclassical) lines correspond to 
$\epsilon_{r1}=\epsilon_{r2}=1$; 
Blue dashed (RPA) and blue dotted (semiclassical) lines correspond to 
$\epsilon_{r1}=4$ and $\epsilon_{r2}=1$. The green (lower) and rose (upper) 
shaded areas represent regimes of intraband and interband excitations, 
respectively.
}
\label{figdisp}
\end{figure*}

Here we consider TM modes in geometry depicted in Fig. 2(a), where graphene is 
surrounded with dielectrics of constants $\epsilon_{r1}$ and $\epsilon_{r2}$. 
Throughout the paper, for definiteness we use $\epsilon_{r1}=4$ corresponding 
to SiO$_2$ substrate, and $\epsilon_{r2}=1$ for air on top of graphene, 
which corresponds to a typical experimental setup. TM modes are found by 
assuming that the electric field has the form
\begin{align}
& E_z=Ae^{iqz-Q_1x},E_y=0,E_x=Be^{iqz-Q_1x},\ \mbox{for $x>0$}, \nonumber \\
& E_z=Ce^{iqz+Q_2x},E_y=0,E_x=De^{iqz+Q_2x},\ \mbox{for $x<0$}.
\label{ansatz}
\end{align}
After inserting this ansatz into Maxwell's equations and matching the boundary 
conditions [which include the conductance of the 2D graphene layer, $\sigma(\omega,q)$], 
we obtain the dispersion relation for TM modes:
\begin{equation}
\frac{\epsilon_{r1}}{\sqrt{q^2-\frac{\epsilon_{r1} \omega^2}{c^2}}}+
\frac{\epsilon_{r2}}{\sqrt{q^2-\frac{\epsilon_{r2} \omega^2}{c^2}}}=
-\frac{\sigma(\omega,q)i}{\omega\epsilon_0}
\label{disp1}
\end{equation}
By explicitly writing the dependence of the conductivity on the wave vector $q$ 
we allow for the possibility of nonlocal effects, where the mean free path of 
electrons can be smaller than $q^{-1}$ \cite{AM}. Throughout this work we 
consider the nonretarded regime ($q\gg \omega/c$), so equation (\ref{disp1}) simplifies to
\begin{equation}
q\approx Q_1\approx Q_2\approx 
\epsilon_0 \frac{\epsilon_{r1}+\epsilon_{r2}}{2}
\frac{2i\omega}{\sigma(\omega,q)}.
\label{disp2}
\end{equation}
Note that a small wavelength (large $q$) leads to a high transversal localization 
of the modes, which are also accompanied by a collective surface charge 
oscillation, similar to SPs in metals; however, it should be understood that, 
in contrast to SPs, here we deal with 2D collective excitations, i.e. plasmons. 
We note that even though field profiles 
of plasmons in graphene and SPs in metals look the same, these two systems are
qualitatively different since electrons in graphene are essentially frozen 
in the transverse dimension \cite{Backes1992}. This fact and the differences 
in electronic dispersions (linear Dirac cones vs. usual parabolic) lead 
to qualitatively different dispersions of TM modes in these two systems 
[see Fig. \ref{figsp}(b) and Fig. \ref{figdisp}(d)]. 
To find dispersion of plasmons in graphene we need 
the conductivity of graphene $\sigma(\omega,q)$, which we now proceed to 
analyze by employing the semiclassical model \cite{AM} (in subsection \ref{subsemi}), 
RPA and number conserving relaxation-time approximation \cite{Mermin1970} (in subsection 
\ref{subRPA}), and by estimating the relaxation-time due to the influence of 
electron-phonon coupling \cite{Park2007} on the optical 
conductivity \cite{Stauber2008} (in subsection \ref{subeph}).

\subsection{Semiclassical model}
\label{subsemi}

For the sake of the clarity of the presentation, we first note that 
by employing a simple semi-classical model for the conductivity 
(see Ref. \cite{AM}), one obtains a Drude-like expression \cite{Hanson2008}: 
\begin{equation}
\sigma(\omega)=\frac{e^2 E_F}{\pi \hbar^2} \frac{i}{\omega+i \tau^{-1}}
\label{sigDrude}
\end{equation}
(the semiclassical conductivity does not depend on $q$). 
Here $\tau$ denotes the relaxation-time (RT), which in a phenomenological 
way takes into account losses due to electron-impurity, electron-defect, 
and electron-phonon scattering. 
Equation (\ref{sigDrude}) is obtained by assuming zero temperature 
$T\approx 0$, which is a good approximation for highly doped 
graphene considered here, since $E_F\gg k_B T$. 
From Eqs. (\ref{disp2}) and (\ref{sigDrude}) it is straightforward to obtain 
plasmon dispersion relation: 
\begin{equation}
q(\omega)=
\frac{\pi \hbar^2 \epsilon_0 (\epsilon_{r1}+\epsilon_{r2})}
{e^2 E_F}(1+\frac{i}{\tau \omega})\omega^2, 
\label{qomDrude}
\end{equation}
as well as losses,
\begin{equation}
\frac{\Re q}{\Im q}=
\omega\tau=
\frac{2\pi c \tau}{\lambda_{air}}. 
\end{equation}
In order to quantify losses one should estimate the relaxation time $\tau$. 
If the frequency $\omega$ is below the interband threshold frequency $\omega_{inter}$, 
and if $\omega<\omega_{Oph}$, then both interband damping and 
plasmon decay via excitation of optical phonons together with an 
electron-hole pair are inactive. 
In this case, the relaxation time can be estimated from DC measurements \cite{Novoselov2004,Geim2007}, 
i.e., it can be identified with DC relaxation 
time which arises mainly from impurities (see Refs. \cite{Novoselov2004,Geim2007}). 
It is reasonable to expect that 
impurity related relaxation time will not display large frequency dependence. 
In order to gain insight into the losses by using this line of reasoning let us 
assume that the doping level is given by $E_F=0.64$~eV (corresponding to 
electron concentration of $n=3\times 10^{13}$~cm$^{-2}$); 
the relaxation time corresponds to DC mobility $\mu=10000$~cm$^2$/Vs 
measured in Ref. \cite{Novoselov2004}: 
$\tau_{DC}=\mu\hbar\sqrt{n\pi}/e v_F=6.4 \times 10^{-13}$s. 
As an example, for the frequency $\hbar\omega=0.155$~eV 
($\lambda_{air}=8\, \mu$m), the semiclassical model yields $\Re q/\Im q \approx 151$ 
for losses and $\lambda_{air}/\lambda_{p}\approx 42$ for wave localization. 
Note that both of these numbers are quite favorable compared to conventional SPs
[e.g., see Fig. \ref{figsp}(c)].
It will be shown in the sequel that for the doping value $E_F=0.64$~eV this
frequency is below the interband loss threshold, and it is evidently also smaller
than the optical phonon loss threshold $\hbar\omega_{Oph}\approx 0.2$~eV, so both of 
these loss mechanisms can indeed be neglected. 

\begin{figure*}[!ht]
\centerline{
%\mbox{\includegraphics[width=0.85\textwidth]{./Figures/Figure3.eps}}
%\mbox{\includegraphics[width=0.85\textwidth]{./Figures/Figure3pdf.eps}}
%\mbox{\includegraphics[width=0.85\textwidth]{./Shito/Figure3.eps}}
\mbox{\includegraphics[width=0.85\textwidth]{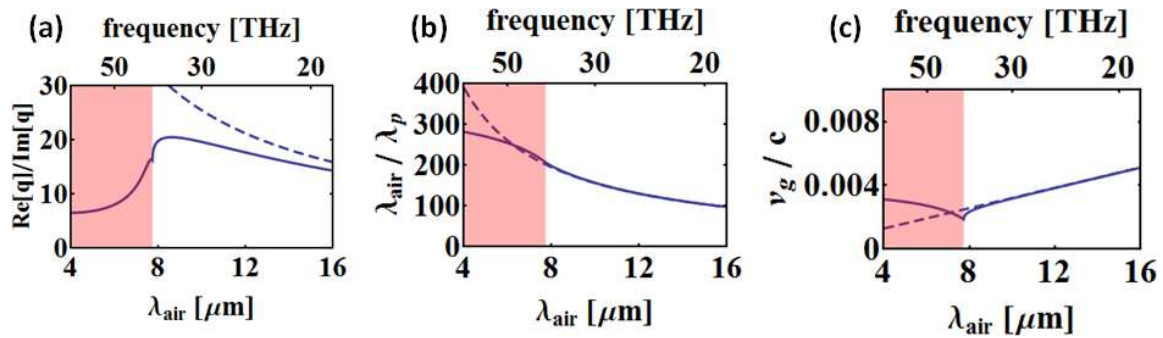}}
}
\caption{
(color online) 
Properties of plasmons in doped graphene. Solid-lines are obtained with the 
number-conserving RPA calculation, and the dashed lines with the semiclassical 
approach. Losses (a), field localization (wave "shrinkage") (b), and group 
velocity (c) for doping $E_F=0.135$~eV, and relaxation time 
$\tau=1.35\times 10^{-13}$~s, which corresponds to the mobility of $10000$~cm$^2/$Vs.
The upper scale in all figures is frequency $\nu=\omega/2\pi$, 
whereas the rose shaded areas denote the region of high interband losses. 
}
\label{figloss}
\end{figure*}

\subsection{RPA and relaxation-time approximation}
\label{subRPA}

In order to take the interband losses into account, we use the self-consistent 
linear response theory, also known as the random-phase approximation (RPA) \cite{AM}, 
together with the relaxation-time (finite $\tau$) approximation introduced by Mermin 
\cite{Mermin1970}. Both of these approaches, 
that is, the collisionless RPA ($\tau\rightarrow \infty$) 
\cite{Wunsch2006,Hwang2007}, and the RPA-RT approximation (finite $\tau$) \cite{Rana2008}, 
have been applied to study graphene. 
In the $\tau\rightarrow \infty$ case, the RPA 2D polarizability of graphene 
is given by \cite{Hwang2007}:
\begin{equation}
\chi(q,\omega) = \frac{e^2}{q^2}\Pi(q,\omega),
\label{chiRPA}
\end{equation}
where
\begin{align}
\Pi(q,\omega) & = \frac{4}{\Omega}\sum_{{\bf k},n_1,n_2}
  \frac{f(E_{n_2,{\bf k}+{\bf q}})-f(E_{n_1,{\bf k}})}
       {\hbar \omega+E_{n_1,{\bf k}}-E_{n_2,{\bf k}+{\bf q}}} \nonumber \\
 & \times |\langle n_1,{\bf k}|e^{-i{\bf q}\cdot {\bf r}}|n_2,{\bf k}+{\bf q} \rangle|^2.
\label{PiRPA}
\end{align}
Here $f(E)=(e^{(E-E_F)/k_BT }+1)^{-1}$ is the Fermi distribution function,
$E_F$ is the Fermi energy and factor 4 stands for 2 spin and 2 valley degeneracies. 
Note that in Eq. (\ref{chiRPA}) $\omega$ is given an infinitesimally small 
imaginary part which leads to the famous Landau damping; that is, 
plasmons can decay by exciting an electron-hole pair 
(interband and intraband scattering) as illustrated in Fig. \ref{figdisp}(c). 
The effects of other types of scattering (impurities, phonons) can be 
accounted for by using the relaxation-time $\tau$ as a parameter 
within the RPA-RT approach \cite{Mermin1970}, which takes into account conservation 
of local electron number. Within this approximation the 2D polarizability is 
\begin{equation}
\chi_{\tau}(q,\omega) =\frac{(1+i/\omega\tau)\chi(q,\omega+i/\tau)}
{1+(i/\omega\tau) \chi(q,\omega+i/\tau)/\chi(q,0)}. 
\label{chitau}
\end{equation}
The 2D dielectric function and conductivity are respectively given by (see 
\cite{Stern1967}):
\begin{equation}
\epsilon_{RPA}(q,\omega)=\frac{\epsilon_{r1}+\epsilon_{r2}}{2}+
\frac{q}{2\epsilon_0} \chi_{\tau}(q,\omega),
\label{epsRPA}
\end{equation}
and 
\begin{equation}
\sigma_{RPA}(q,\omega)=-i\omega \chi_{\tau}(q,\omega).
\label{sigmaRPA}
\end{equation}
We note here that throughout the text only $\pi$–--bands are taken into 
consideration; it is known that in graphite, higher $\sigma$–--bands give 
rise to a small background dielectric constant \cite{Taft1965} at low energies, 
which is straightforward to implement in the formalism. Using Eqs. (\ref{disp2}) 
and (\ref{sigmaRPA}) we obtain that the properties of plasmons (i.e., dispersion, 
wave localization and losses) can be calculated by solving 
\begin{equation}
\epsilon_{RPA}(q,\omega)=0,
\label{eps=0}
\end{equation}
with complex wave vector $q=q_1+iq_2$. The calculation is simplified by 
linearizing Eq. (\ref{eps=0}) in terms of small $q_2/q_1$, to obtain,
\begin{equation}
\frac{\epsilon_{r1}+\epsilon_{r2}}{2}+
\frac{e^2}{2\epsilon_0 q_1} \Re [\Pi(q_1,\omega)]=0,
\label{dispfin}
\end{equation}
for the plasmon dispersion, and 
\begin{widetext}
\begin{equation}
q_2=\frac{\Im [\Pi(q_1,\omega)]+\frac{1}{\tau}
\frac{\partial}{\partial\omega} \Re [\Pi(q_1,\omega)]
+\frac{1}{\omega\tau} \Re [\Pi(q_1,\omega) (1-\Pi(q_1,\omega))/\Pi(q_1,0)]}
{\frac{1}{q_1}\Re [\Pi(q_1,\omega)]-\frac{\partial}{\partial q_1}\Re [\Pi(q_1,\omega)]}
\label{q2}
\end{equation}
\end{widetext}
yielding losses. 
Note that in the lowest order the dispersion relation 
(and consequently $\lambda_{air}/\lambda_p$ and the group velocity 
$v_g$) does not depend on $\tau$. 
This linearization is valid when $q_2 \ll q_1$; as the plasmon 
losses increase, e.g., after entering the interband regime [the rose area in Fig. 
\ref{figdisp}(d)], results from Eqs. (\ref{dispfin}) and (\ref{q2}) should be 
regarded as only qualitative. The characteristic shape of the plasmon 
dispersion is shown in Fig. \ref{figdisp}(d). 
Note that the semi-classical model and the RPA model agree well 
if the system is sufficiently below the interband threshold
[for small $q$, $\omega(q)\sim \sqrt{q}$ as in Eq. (\ref{qomDrude})]. 
By comparing Figs. \ref{figdisp}(d) and \ref{figsp}(b) we see that the dispersion 
for SPs on silver-dielectric surface qualitatively differs from the plasmon 
dispersion in graphene \cite{Backes1992}.
While SPs' dispersion relation approaches an asymptote ($\omega\rightarrow\omega_{SP}$) 
for large $q$ values [Eq. (\ref{convP})], graphene plasmon relation gives 
$\omega(q)$ which continuously increases [Fig. \ref{figdisp}(d)].

Theoretically predicted plasmon losses $\Re q/\Im q$ and wave 
localization $\lambda_{air}/\lambda_{p}$ are illustrated in Fig. 
\ref{figloss} for doping level $E_F=0.135$~eV and relaxation time 
$\tau=1.35\times 10^{-13}$~s. 
We observe that for this particular doping level, 
for wavelengths smaller than $\lambda_{inter}\approx 7.7\, \mu$m, 
the system is in the regime of high interband losses (rose shaded region). 
Below the interband threshold, both losses and wave localization obtained by 
employing RPA-RT approach are quite well described by the previously 
obtained semiclassical formulae. 
Since the frequencies below the interband threshold are (for the assumed 
doping level) also below the optical phonon frequency, the relaxation time 
can be estimated from DC measurements. 

At this point we also note that in all our calculations we have 
neglected the finite temperature effects, i.e., $T\approx 0$. 
To justify this, we note that for doping values utilized in this paper 
the Fermi energies are $0.135$~eV$\approx 5.2k_BT_{r}$ ($n=1.35\times 10^{12}$~cm$^{-2}$) 
and $0.64$~eV$\approx 25k_BT_{r}$ ($n=3\times 10^{13}$~cm$^{-2}$) for room 
temperature $T_{r}=300$~K.
The effect of finite temperature is to slightly smear the sharpness of the 
interband threshold, but only in the vicinity ($\sim k_BT_{r}$) of the threshold.

By increasing the doping, $E_F$ increases, and the region of interband 
plasmonic losses moves towards higher frequencies (smaller wavelengths). 
However, by increasing the doping, 
the interband threshold frequency will eventually become larger than graphene's optical 
phonon frequency $\omega_{Oph}$: there will exist an interval of 
frequencies, $\omega_{Oph}<\omega<\omega_{inter}$, where it is kinematically 
possible for the photon of frequency $\omega$ to excite an 
electron-hole pair together with emission of an optical phonon. 
This second order process can reduce the relaxation time estimated
from DC measurements and should be taken into account, as we show in the 
following subsection.

\begin{figure*}[!ht]
\centerline{
%\mbox{\includegraphics[width=0.65\textwidth]{./Figures/Figure4.eps}}
%\mbox{\includegraphics[width=0.65\textwidth]{./Figures/Figure4pdf.eps}}
%\mbox{\includegraphics[width=0.85\textwidth]{./Shito/Figure4.eps}}
\mbox{\includegraphics[width=0.75\textwidth]{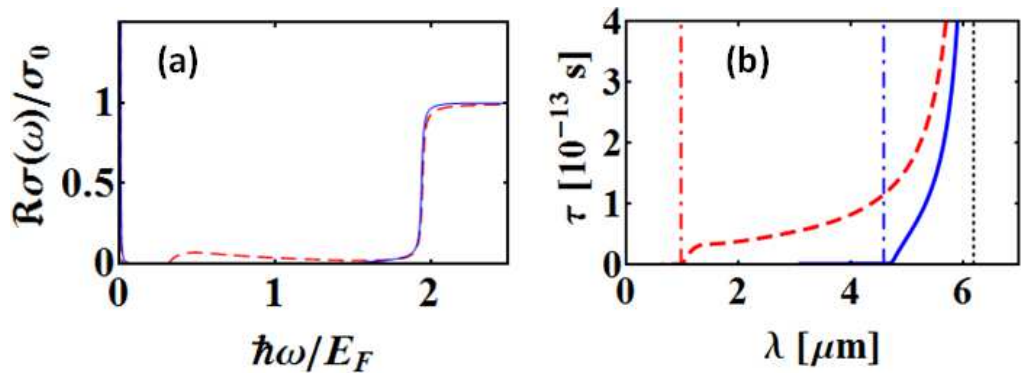}}
}
\caption{
(color online) (a) The real part of the conductivity in units of $\sigma_0=\pi e^2/2h$ 
in dependence of frequency $\hbar\omega/E_F$, and (b) the corresponding relaxation time 
as a function of wavelength. The contribution to $\Re\sigma(\omega)$ from impurities 
is chosen to be negligible. The displayed graphs correspond to two different values 
of doping which yield $E_F=0.135$ eV (solid blue line), and 
$E_F=0.640$ eV (dashed red line). The position of the optical phonon frequency 
$\hbar\omega_{Oph}\approx0.2$ eV is depicted by the dotted vertical line in (b); 
dot-dashed lines depict the values of wavelengths corresponding to $2E_F$, that is, 
the interband threshold value (for $q=0$) for the two doping concentrations. 
}
\label{sigtau}
\end{figure*}

\subsection{Losses due to optical phonons}
\label{subeph}

In what follows, we estimate and discuss the relaxation time due to the 
electron-phonon coupling. 
This can be done by using the Kubo formula which has been utilized in 
Ref. \cite{Stauber2008} to calculate the real part of the 
optical conductivity, $\Re\sigma(\omega,q=0)$. The calculation of conductivity 
$\Re\sigma(\omega,0)$ involves the electron self-energy 
$\Sigma(E)$, whose imaginary part expresses the width of a 
state with energy $E$, whereas the real part corresponds to the energy shift. 
Let us assume that the electron self-energy stems from the 
electron-phonon coupling and impurities, 
\begin{equation}
\Sigma(E)=\Sigma_{e-ph}(E)+\Sigma_{imp}(E). 
\end{equation}
For $\Sigma_{e-ph}$ we utilize a simple yet fairly accurate model 
derived in Ref. \cite{Park2007}: If $|E-E_F|>\hbar\omega_{Oph}$, 
then 
\begin{equation}
\Im\Sigma_{e-ph}(E)=\gamma |E-\mbox{sgn}(E-E_F)\hbar \omega_{Oph}|,
\end{equation}
while elsewhere $\Im\Sigma_{e-ph}(E)=0$; the dimensionless constant 
$\gamma=18.3\times 10^{-3}$ \cite{Park2007} is proportional to the 
square of the electron-phonon matrix element \cite{Park2007}, i.e., 
the electron-phonon coupling coefficient.
In order to mimic impurities, we will assume that $\Im\Sigma_{imp}(E)$ 
is a constant (whose value can be estimated from DC measurements). 
The real parts of the self-energies are calculated 
by employing the Kramers-Kr\" onig relations. In all our calculations 
the cut-off energy is taken to be $8.4$~eV, which corresponds to the 
cut-off wavevector $k_{c}=\pi/a$, where $a=2.46$~\AA. 
By employing these self-energies we calculate the conductivity 
$\Re\sigma(\omega,q=0)$, from which we estimate the relaxation time 
by using Eq. (\ref{sigDrude}), i.e., 
\begin{equation}
\tau(\omega)\approx 
\frac{e^2 E_F}{\pi \hbar^2 \omega^2}
\frac{1}{\Re \sigma (\omega,0)}
\label{taufin}
\end{equation}
for the region below the interband threshold; in deriving (\ref{taufin}) 
we have assumed $\tau\omega\gg 1$.

Figure \ref{sigtau} plots the real part of the conductivity and the relaxation time 
for two values of doping: $E_F=0.135$~eV ($n=1.35\times 10^{12}$~cm$^{-2}$, 
solid line) and $E_F=0.64$~eV ($n=3\times 10^{13}$~cm$^{-2}$, dashed line). 
In order to isolate the influence of the electron-phonon coupling 
on the conductivity and plasmon losses, the contribution from impurities 
is assumed to be very small: $\Im\Sigma_{imp}(E)=10^{-6}$~eV. 
The real part of the conductivity has a universal value $\sigma_0=\pi e^2/2h$ 
above the interband threshold value $\hbar\omega=2E_F$ (for $q=0$), e.g., see 
\cite{Basov2008,Nair2008,Mak2008}. 
We clearly see that the relaxation time is not affected by the 
electron-phonon coupling for frequencies below $\omega_{Oph}$, that is, 
we conclude that scattering from impurities and defects is a dominant decay 
mechanism for $\omega<\omega_{Oph}$ (assuming we operate below the interband threshold). 
However, for $\omega>\omega_{Oph}$, the relaxation times in Fig. \ref{sigtau} 
are on the order of $10^{-14}-10^{-13}$~s, 
indicating that optical phonons are an important decay mechanism. 

It should be emphasized that the exact calculated values should be 
taken with some reservation for the following reason: strictly speaking, one should 
calculate the relaxation times $\tau(\omega,q)$ along the plasmon 
dispersion curve given by Eq. (\ref{dispfin}); namely the matrix elements
which enter the calculation depend on $q$, whereas the phase space 
available for the excitations also differ for $q=0$ and $q>0$. 
Moreover, the exact value of the matrix element for electron phonon coupling 
is still a matter of debate in the community (e.g., see Ref. \cite{McChesney2008}). 
Therefore, the actual values for plasmon losses could be somewhat different 
for $\omega>\omega_{Oph}$. Nevertheless, fairly small values of relaxation times presented in 
Fig. \ref{sigtau} for $\omega>\omega_{Oph}$ indicate that emission of 
an optical phonon together with an electron-hole pair is an important decay 
mechanism in this regime. Precise calculations for $q>0$ and $\omega>\omega_{Oph}$ 
are a topic for a future paper. 

\begin{figure*}[!ht]
\centerline{
%\mbox{\includegraphics[width=0.85\textwidth]{./Figures/Figure5.eps}}
%\mbox{\includegraphics[width=0.85\textwidth]{./Figures/Figure5pdf.eps}}
%\mbox{\includegraphics[width=0.85\textwidth]{./Shito/Figure5.eps}}
\mbox{\includegraphics[width=0.85\textwidth]{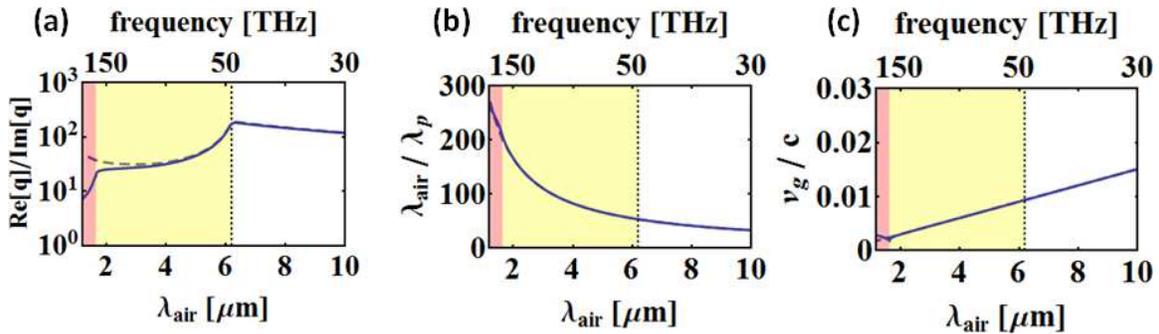}}
}
\caption{
(color online) 
Properties of plasmons in doped graphene. Solid-lines are obtained with the 
number-conserving RPA calculation, and the dashed lines with the semiclassical 
approach. Losses (a), field localization (wave "shrinkage") (b), and group 
velocity (c) for doping $E_F=0.64$~eV; losses are calculated by 
using the relaxation time $\tau^{-1}=\tau^{-1}_{DC}+\tau^{-1}_{e-ph}$, where 
$\tau_{DC}=6.4\times 10^{-13}$~s, and $\tau_{e-ph}$ is the relaxation time 
from the electron-phonon coupling for the given parameters. 
In the white regions (right regions in all panels), losses are 
determined by $\tau_{DC}$. 
In the yellow shaded regions (central regions in all panels), losses are 
determined by the optical phonon emission, i.e., $\tau_{e-ph}$. 
The rose shaded areas (left region in all panels) denote the region 
of high interband losses.
Dotted vertical lines correspond to the optical phonon frequency 
$\omega_{Oph}\approx 0.2$~eV. The upper scale in all figures is 
frequency $\nu=\omega/2\pi$. 
 See text for details. 
}
\label{figloss5}
\end{figure*}

Plasmonic losses and wave localization calculated from the RPA-RT approximation 
are illustrated in Fig. \ref{figloss5} for doping level $E_F=0.64$~eV
and the relaxation time $\tau$ given by $\tau^{-1}=\tau^{-1}_{DC}+\tau^{-1}_{e-ph}$, where 
$\tau_{DC}=6.4\times 10^{-13}$~s (mobility $10000$~cm$^2/$Vs), 
whereas $\tau_{e-ph}$ is frequency dependent and corresponds to electron-phonon 
coupling assuming very clean samples 
[see dashed line in Fig. \ref{sigtau}(b)]. 
Interband losses [left (rose shaded) regions in all panels] are active for wavelengths smaller 
than $\lambda_{inter}\approx 1.7\,\mu$m. In the frequency interval 
$\omega_{inter}>\omega>\omega_{Oph}$ [central (yellow shaded) regions in all panels], 
the decay mechanism via electron phonon coupling determines the loss rate, 
i.e., $\tau\approx \tau_{e-ph}$. 
For $\omega<\omega_{Oph}$ [right (white) regions in all panels], the DC 
relaxation time $\tau_{DC}$ can be used to estimate plasmon losses.

It should be noted that the mobility of $10000$~cm$^2/$Vs could be improved, 
likely even up to mobility $100000$~cm$^2/$Vs \cite{Geim2007}, thereby further improving 
plasmon propagation lengths for frequencies below the optical phonon frequency. 
However, for these larger mobilities the calculation of losses should also include 
in more details the frequency dependent contribution to the relaxation time from 
acoustic phonons (this decay channel is open at all frequencies); 
such a calculation would not affect losses for $\omega>\omega_{Oph}$ where 
optical phonons are dominant.

\section{Conclusion and Outlook}
\label{sec:conc}

In conclusion, we have used RPA and number-conserving relaxation-time approximation 
with experimentally available input parameters, and theoretical estimates for the 
relaxation-time utilizing electron-phonon coupling, to study plasmons and their 
losses in doped graphene. 
We have shown that for sufficiently large doping values
high wave localization and low losses are simultaneously 
possible for frequencies below that of the optical phonon branch 
$\omega<\omega_{Oph}$ (i.e., $E_{plasmon}<0.2$~eV). 
For sufficiently large doping values, 
there is an interval of frequencies above $\omega_{Oph}$ and below 
interband threshold, where an important decay mechanism for plasmons 
is excitation of an electron-hole pair together with an optical 
phonon (for $\omega<\omega_{Oph}$ this decay channel is inactive);
the relaxation times for this channel were estimated and discussed. 
We point out that further more precise calculations of plasmon relaxation 
times should include coupling to the substrate (e.g., coupling to surface-plasmon 
polaritons of the substrate), a more precise shape of the phonon dispersion 
curves  \cite{phonondisp}, and dependence of the relaxation time via 
electron-phonon coupling on $q>0$ (see subsection \ref{subeph}). 

The main results, shown in Figures \ref{figloss} and \ref{figloss5} point out 
some intriguing opportunities offered by plasmons in graphene for the field of 
nano-photonics and metamaterials in infrared (i.e. for $\omega<\omega_{Oph}$). 
For example, we can see in those figures that high field localization and enhancement  
$\lambda_{air}/\lambda_p\sim 200$ [see Figure \ref{figloss}(b)] are possible 
(resulting in $\lambda_p<50$~nm), while plasmons of this kind could have propagation 
loss-lengths as long as $\sim 10\lambda_p$ [see Fig. \ref{figloss5}(a)];
these values (albeit at different frequencies) are substantially 
more favorable than the corresponding values for 
conventional SPs, for example, for SPs at the Ag/Si interface 
$\lambda_{air}/\lambda_p\sim 20$, whereas propagation lengths are only 
$\sim 0.1\lambda_{sp}$ [see Fig. \ref{figsp}(c)]. 
Another interesting feature of plasmons in graphene is that, similar to usual 
SP-systems \cite{Aristeidis2005}, wave localization is followed by a group 
velocity decrease; the group velocities can be of the order 
$v_g=10^{-3}-10^{-2}$c, and the group velocity can be low over a wide frequency 
range, as depicted in Figs. \ref{figloss}(c) and \ref{figloss5}(c). This is of interest 
for possible implementation of novel nonlinear optical devices in graphene, 
since it is known that small group velocities can lead to savings in both the 
device length and the operational power \cite{Soljacic2004}; the latter would also be reduced 
because of the large transversal field localization of the plasmon modes.

%=======================================================================================

\acknowledgments
We would like to thank Leonid Levitov, J.D. Joannopoulos, Pablo Jarillo-Herrero, 
Tony Heinz, Vladimir Shalaev, Thomas Ebbesen, Ivan Kup\v ci\' c, Antonio \v Siber, 
and Branko Gumhalter, Tobias Stauber, and Ivo Batisti\' c, for many helpful comments. 
This work was supported in part by the Croatian Ministry of Science, 
Grant No. 119-0000000-1015. This work was also supported in part by the 
MRSEC program of National Science Foundation under award number DMR-0819762, 
in part by the U.S. Army Research Office through the institute of soldier 
nanotechnologies under contract No. W911NF-07-D-0004, and also in part by the U.S. 
Department of Energy office of science, office of basic energy sciences, 
through S3TEC, which is a DOE energy frontier research center.

%=======================================================================================


\begin{thebibliography}{99}
\bibitem{phonondisp}
For simplicity here we assume Einstein model for optical phonons, and assume that 
they form a branch at $0.2$~eV. More precisely in graphene there are several 
types of optical phonons interacting with electrons, e.g., Zone Center phonons 
at $\hbar\Omega_{\Gamma}=0.196$~eV, and Zone Boundary phonons 
at $\hbar\Omega_{K}=0.161$~eV.
\end{thebibliography}
\end{document}